\documentclass[fleqn,10pt]{wlscirep}
\usepackage[utf8]{inputenc}
\usepackage[T1]{fontenc}
\usepackage{verbatim}
\title{Barium Chemosensors with Dry-Phase Fluorescence for Neutrinoless Double Beta Decay}

\author[1,2*]{P. Thapa}
\author[3]{I. Arnquist}
\author[1]{N. Byrnes}
\author[2]{A. A. Denisenko}
\author[2]{F. W. Foss Jr.}
\author[1]{B. J. P. Jones}
\author[1]{A. D. McDonald}
\author[1]{D. R. Nygren}
\author[1]{K. Woodruff}
\affil[1]{Department of Physics, University of Texas at Arlington, Arlington, TX 76019, USA}
\affil[2]{Department of Chemistry and Biochemistry, University of Texas at Arlington, Arlington, TX 76019, USA}
\affil[3]{Pacific Northwest National Laboratory (PNNL), Richland, WA 99352, USA}
\affil[*]{pawan.thapa@uta.edu}

\begin{abstract}
The nature of the neutrino is one of the major open questions in experimental nuclear and particle physics.  The most sensitive known method to establish the Majorana nature of the neutrino is detection of the ultra-rare process of neutrinoless double beta decay.  However, identification of one or a handful of decay events within a large mass of candidate isotope, without obfuscation by backgrounds is a formidable experimental challenge.  One hypothetical method for achieving ultra-low-background neutrinoless double beta decay sensitivity is the detection of single $^{136}$Ba ions produced in the decay of $^{136}$Xe (``barium tagging'').  To implement such a method, a single-ion-sensitive barium detector must be developed and demonstrated in bulk liquid or dry gaseous xenon.  This paper reports on the development of two families of dry-phase barium chemosensor molecules for use in high pressure xenon gas detectors, synthesized specifically for this purpose.  One particularly promising candidate, an anthracene substituted aza-18-crown-6 ether, is shown to respond in the dry phase with almost no intrinsic background from the unchelated state, and to be amenable to barium sensing through fluorescence microscopy.  This interdisciplinary advance, paired with earlier work demonstrating sensitivity to single barium ions in solution, opens a new path toward single ion detection in high pressure xenon gas.

\end{abstract}
\begin{document}

\flushbottom
\maketitle
%
%
\thispagestyle{empty}

\section{Introduction}

The search for neutrinoless double beta decay ($0\nu\beta\beta$) is a major focal point of experimental nuclear physics worldwide. If and only if the neutrino is a Majorana fermion (a particle that is its own antiparticle), the lepton-number violating nuclear decay $^{N}_{Z}X\rightarrow ^{N}_{Z-2}X+2e$ may take place.  An observation of this decay would simultaneously prove that lepton number is not conserved in nature; demonstrate the existence of physics beyond the standard model responsible for the lightness of neutrino masses~\cite{Chang:1985en,minkowski1977mu,gell1979ramond,yanagida1979proceedings,mohapatra1981neutrino};  and lend weight for leptogenesis as the mechanism for introducing matter-antimatter asymmetry into an initially symmetrical Universe~\cite{Fukugita:1986hr}. The search for $0\nu\beta\beta$ has spurred the development of many sensitive detection techniques.

The primary challenges involved in the search for $0\nu\beta\beta$ derive from the fact that this process, if it occurs at all, is extremely slow. Present experimental lower limits on $0\nu\beta\beta$ lifetime sit at $1.2\times 10^{26}$ years~\cite{KamLAND-Zen:2016pfg}. Well-motivated theoretical models involving light neutrino exchange predict that $0\nu\beta\beta$ could be observed with any lifetime beyond this limit~\cite{GomezCadenas:2011it,Bilenky:2012qi,Vergados:2012xy}, given present knowledge of neutrino masses and mixing angles. Observing such a rare process above experimental background from sources such as detector material gamma rays~\cite{Martin-Albo:2015rhw}, dissolved radioisotopes~\cite{Novella:2018ewv,Rupp:2017zcy}, neutron captures~\cite{Carson:2004cb}, and others is a formidable experimental challenge, requiring deep underground detectors, exquisite radio-purity~\cite{Alvarez:2012as,Alvarez:2014kvs,Cebrian:2017jzb}, and extremely selective signal identification and background rejection methods.

The target for the next generation of experiments is to deploy ton- to multi-ton-scale detectors with background rates of less than 0.1 counts per ton per year in the experimental region of interest (ROI).  At the time of writing, the best demonstrated background rate from any technology has been demonstrated by germanium diodes, and is around four in these units~\cite{Agostini:2018tnm,Aalseth:2017btx}. 

One isotope that has drawn particular attention as a suitable candidate for building very large, low background $0\nu\beta\beta$ experiments is $^{136}$Xe, which could decay via $0\nu\beta\beta$ to $^{136}$Ba+2e.  Technologies have been developed to search for $0\nu\beta\beta$ of $^{136}$Xe in both liquid (LXe)~\cite{Albert:2017owj} and gas (GXe)~\cite{Renner:2018ttw} phases, and dissolved in liquid scintillator (LSXe)~\cite{KamLAND-Zen:2016pfg}, with the latter providing the present strongest limit.  The lowest background rate from a running $0\nu\beta\beta$ search in $^{136}$Xe is from the EXO-200 experiment using LXe, and is $\sim 144$ counts per ton per year in the ROI~\cite{Albert:2017owj}. The NEXT program is presently constructing a 100~kg high pressure GXe (HPGXe) detector called NEXT-100 that projects a background index of between 5 and 10 counts per ton per year in the ROI~\cite{Martin-Albo:2015rhw}. Despite impressive technological advances from all techniques, an effectively background-free $0\nu\beta\beta$ technology for the ton-scale and beyond remains undemonstrated.

As recognized 17 years ago~\cite{Moe:1991ik}, in experiments using the isotope $^{136}$Xe, efficient and selective detection of the daughter nucleus $^{136}$Ba$^{2+}$,
accompanied by electron energy measurements of precision better than $\sim$2\%
FWHM to reject the two-neutrino double beta decay background, could provide an effectively background free approach to search for $0\nu\beta\beta$.  Recently, single-atom or single-ion-sensitive detection methods for detection of barium have been demonstrated, emerging from various disciplines in physics and chemistry~\cite{McDonald:2017izm,Chambers:2018srx}. As yet, no sensor capable of direct deployment within the working medium of a running detector has been produced.

A key consideration differentiating barium tagging approaches in LXe and GXe experiments is the charge state of the daughter nucleus. Barium from double beta decay is born in a highly ionized state Ba$^{N+}$ that quickly captures electrons from neutral Xe until further capture is energetically disfavored, stopping at Ba$^{2+}$.
In LXe, recombination with locally thermalized electrons causes further neutralization, leading to an ensemble of ionic and atomic species\cite{Albert:2015vma} including Ba and Ba$^{+}$. The lack of recombination in GXe, on the other hand, implies that Ba$^{2+}$ will be the predominant outcome~\cite{Novella:2018ewv}.  Optimal technologies are likely to be distinct for detection in these two cases.

Barium tagging approaches for LXe \cite{Sinclair:2011zz} have previously been focused on atomic fluorescence transitions of the outer electron in Ba$^+$, or in some cases, neutral Ba~\cite{Mong:2014iya}.  Ba$^{2+}$, on the other hand, has a noble-like electron configuration without low-lying fluorescence transitions.
To detect Ba$^{2+}$ using fluorescence techniques it is necessary
to add such transitions artificially. One such method was proposed in~\cite{Elba} and further developed in~\cite{Jones:2016qiq}, using  Single Molecule Fluorescence Imaging (SMFI). SMFI and super-resolution techniques, resulting in the 2014 Nobel prize in chemistry, are widely used in biology for sensing Ca$^{2+}$ ions at the single ion level~\cite{stuurman2006imaging,fish2009total,Thomas2000,Lu2007,Oliver2000,nakahara2005new}.
In SMFI detection of cations, a suitable fluorophore-containing molecule is employed that is nearly non-fluorescent
in isolation but becomes highly fluorescent upon chelation of a target
ion (Fig.~\ref{fig:BindingCartoon}). The fluorescence enabled upon binding can be observed by probing
with an excitation laser and collecting the fluorescence light using single-photon-sensitive EM-CCD cameras.  The range of applicability of SMFI dyes is wide, including in solution and inside living cells.  This work focuses on the adaptation of SMFI for sensing of dications in dry gaseous environments at the solid-gas interface.

\section{Development of SMFI dyes for dry-phase barium ion sensing}

\begin{figure}[t]
\begin{centering}
\includegraphics[width=0.99\columnwidth]{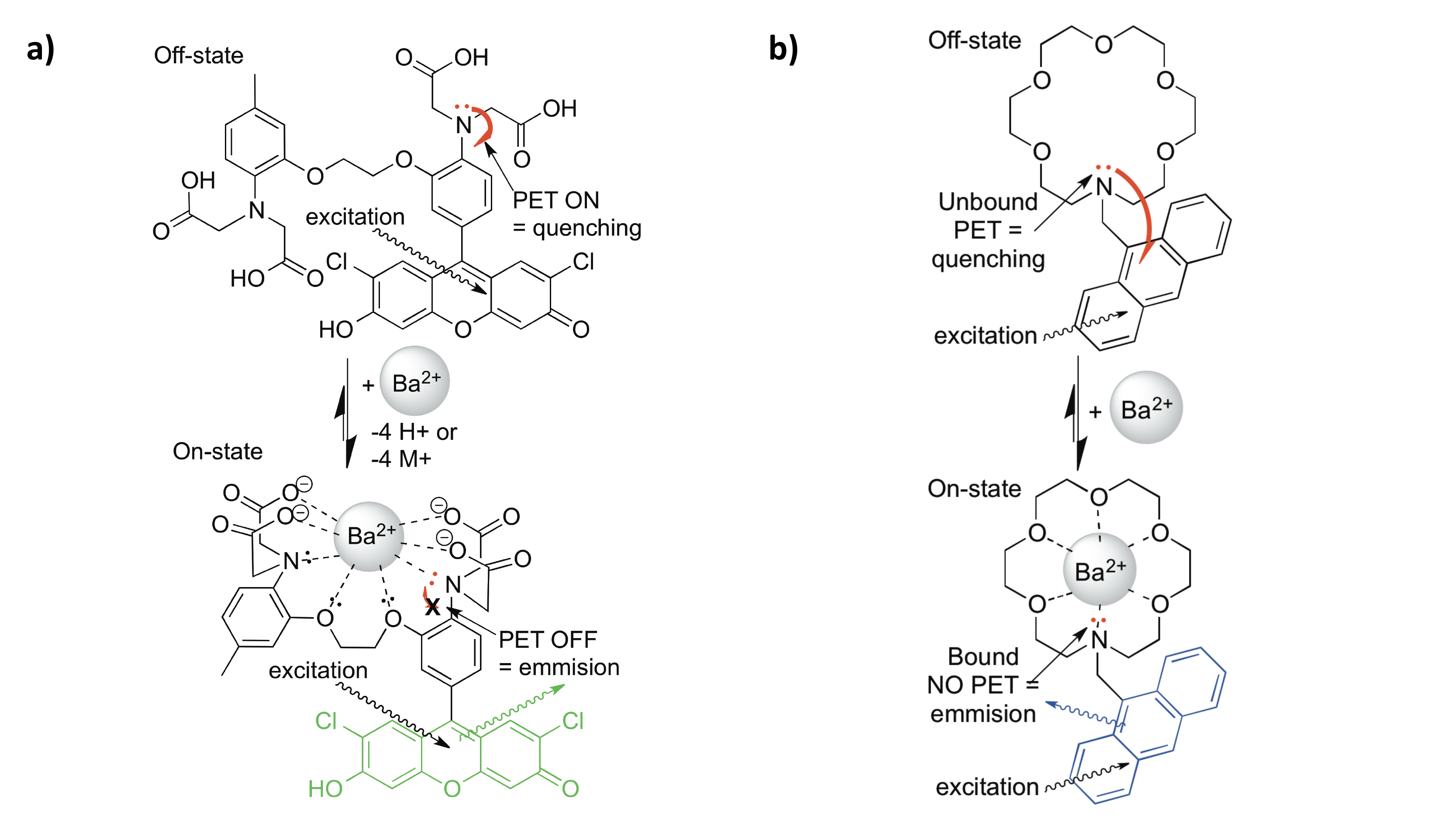}
\par\end{centering}
\caption{Cartoon showing binding and turn-on fluorescence in dyes a) FLUO-3, used in~\cite{Jones:2016qiq,McDonald:2017izm} and b) 18c6-an, developed for this work.\label{fig:BindingCartoon}}
\end{figure}

A cross-disciplinary effort has been underway since 2015 to develop barium tagging techniques for high pressure xenon gas detectors using single molecule fluorescent imaging.  In  exploratory work, we demonstrated
that commercially available calcium chemosensors FLUO-3 and FLUO-4 are sensitive probes for Ba$^{2+}$, thus making them potentially promising barium tagging agents~\cite{Jones:2016qiq}. Using
total internal reflection fluorescence microscopy~\cite{axelrod2003} (TIRFM) with FLUO-3, the first single
barium dication fluorescence detection was then demonstrated~\cite{McDonald:2017izm}. Individual ions were spatially resolved with 2 nm super-resolution and identified
with 13~$\sigma$ statistical significance over background via their
sharp photo-bleaching transitions.

FLUO-3 and FLUO-4 are formed from a 1,2-bis(\textit{o}-aminophenoxy)ethane-\textit{N,N,N',N'}-tetraacetic acid (BAPTA)-like receptor that is
covalently bound to a fluorescein-like fluorophore. On-off fluorescence response in the FLUO family arises from quenching of the fluorescein group, presumably by photo-induced electron transfer from
the electron-rich BAPTA species (likely involving the electrons on the adjacent nitrogen atom, see Fig.~\ref{fig:BindingCartoon}, left). When unbound to an
ion, electrons may move freely into the fluorescein group to
quench excited states and inhibit fluorescence. In the presence of Ba$^{2+}$, the electronics of the BAPTA-like receptor are altered, effectively by Lewis acid-base complexation with the nitrogen's lone pair electrons. This inhibits the donation of electrons into the excited fluorescein group, and prevents the flourescence quenching, switching on a fluorescent response. These changes in fluorescent quenching states can be probed by laser excitation at an energy just above the fluorescence transition.

The majority of contemporary SMFI work has been performed in solution, often with molecules in pockets of liquid suspended in a polymer matrix.
For example, our previous work~\cite{McDonald:2017izm} used
FLUO-3 suspended in polyvinyl alcohol to detect
ions from barium perchlorate solution. On the other hand, the xenon gas within $0\nu\beta\beta$ experiments must be free of impurities such as water at the part-per-trillion level, which is achieved through continuous circulation and purification with hot and cold getters~\cite{Monrabal:2018xlr}. Any water-based sensor is therefore inappropriate for this application, and a dry-phase barium chemosensor is required. Our studies of the FLUO family implied they will be unsuitable for barium detection in the dry phase, for two reasons: 1) binding of BAPTA to Ba$^{2+}$ requires deprotonation
of four carboxylic acid groups and reorganization of the various points of binding, not expected
to occur effectively in a dry state (Fig.~\ref{fig:BindingCartoon}). 2) While fluorescein is a bright fluorophore in solution, the fluorescence is suppressed when dried~\cite{Byrnes:2019jxr}. 

Similarly, most ion sensing work in SMFI focuses on calcium detection, due to its relevance as a signalling agent in biological systems.  Some prior work on chemosensors for barium ions has been reported~\cite{Nakahara2004,chaichana2019selective,kondo2011synthesis}, but barium sensing remains a relatively unexplored sub-field.

We have thus initiated an interdisciplinary program to develop and test barium-sensitive molecules for use in the dry phase, to enable barium tagging in high pressure xenon gas.  The requirements for a barium chemosensor suitable for ion tagging in $0\nu\beta\beta$ include 1) relatively long wavelength excitation and emission, suitable for transmission through microscope optics (in practice both excitation and emission longer than ~350 nm); 2) bright fluorescence of the ``on'' (barium-bound) state in the dry phase; 3) a strongly suppressed ``off'' (barium-unbound) state in the dry phase;  4) A sufficient Stokes shift to allow dichroic separation of excitation and emission light; 5)  binding that is strong enough to reliably capture ions but weak enough that residual ions can be washed from the sensor with a competing stronger receptor before installation (likely anything in range of dissociation constants $K_d\sim10\mu$M to few mM range is suitable); 6) ability to chelate ions at the gas-solid interface; 7) stability under a dry noble gas environment. In this work we address points 1 through 5. Points 6 and 7 depend sensitively on the form of the final coating, envisaged here as a self-assembled and covalently tethered monolayer on a thin transparent surface, to be developed and addressed in future work.

In order to identify daughter ions at a sensing region of an $0\nu\beta\beta$ experiment, they must first be transported from the position of the decay to the sensor.  There are various lines of technological development being pursued to address this question. The NEXT collaboration~\cite{Gomez-Cadenas:2019sfa} is undertaking R\&D on RF carpet structures\cite{bollen2011ion} to achieve electric-field driven transport in high pressure xenon gas detectors.  A proposed scheme for barium tagging in liquid xenon also involves RF methods~\cite{Brunner:2014sfa}, in this case to extract from gas ullage into an evacuated region.  This could be coupled to mechanical actuation of a cold finger to extract frozen ions from liquid into the gas~\cite{gornea2011search}.     For a realistic application, the efficiency of the barium tagging process incorporating both transport and detection must be high. A dense monolayer of molecular probes may plausibly exhibit an ion tagging efficiency near one.  Establishing this number will require detailed ion-by-ion characterization in the dry phase, a test we plan to undertake at the University of Texas at Arlington as a future step in this program.

Considering the need for molecular structures that function in dry systems, aza-crown ethers appear to be an ideal candidate for photo-induced electron transfer (PET)\cite{escudero2016revising} modulated turn-on SMFI techniques. They have been successfully used for the capture of various earth and earth alkali metals in both protic and aprotic solutions and their nitrogen can be placed in a fashion to modulate the excited states of various fluorophores (including organic dyes, nanoparticles, and hybrid fluorophores)\cite{hughes2017aza,roy2016amplified,dhenadhayalan2016silicon,spath2010luminescent}. As binding domains for alkali earth metals, aza-crown ethers demonstrate slightly higher $K_d$ values as compared to their all-oxygen crown ether precursors\cite{krakowiak1989synthesis}. Their binding, however, is expected to achieves similar selectivity based on size considerations and the pre-organized (conformationally restrained) ring systems.  (Fig.~\ref{fig:BindingCartoon}) demonstrates the chemical bonding model supported by crystal structures and NMR experiments, depicting binding by cooperative Lewis base donation of lone pair electrons of the crown ether oxygens and nitrogen to cations and other Lewis acidic guest molecules.  Aza-crown ethers may be covalently linked to fluorophore molecules through their nitrogen atom, forming tertiary amines or secondary amides, allowing for divergent synthesis of various classes of fluorescent sensors. Both amine and amide functional groups act as ``switches'' for modulation of light and dark states in tethered fluorophores\cite{wang2014density}-although, these two moieties display different PET donation abilities to acceptor fluorophores and innate stability. Stemming from the structural arguments above, the aprotic and preorganized binding domain of aza-18-crown-6 ether was envisioned for optimal chelation of Ba$^{2+}$. The aza-15-crown-5 ether has a relatively smaller size and in some cases has been shown to chelate Ba\textsuperscript{2+} in a termolecular, sandwich-like complex with a 2:1 ratio of crown-ethers to Ba\textsuperscript{2+}.~\cite{gromov2009n,kondo2011synthesis} Aza-21-crown-7 ether, with its larger ring size is thought to result in weaker/longer bonds to metal ions, which cannot overcome the entropy and flexibility seen in larger ring systems.~\cite{lochman2015role,shinkai1982photoresponsive}.

When considering fluorophores for sensitive function, especially in dry and solid states, it is common to select more rigid aromatic or heteroaromatic fluorophores, which are relatively photostable and undergo less structural reorganization and non-radiative deactivation pathways after excitation~\cite{moerner2003methods}. We chose to investigate pure arenes (all carbon and hydrogen aromatic fluorophores) in this initial investigation of functional SMFI molecules because of their robust nature, ease of synthetic tailoring, and relatively weak intermolecular interactions with barium ions. This paper presents the first family of molecules designed specifically for this purpose, experimentally investigating aza-crown ether binding site of various sizes bound through the crown-ether amine to a methyl aromatic fluorophore, creating a benzylic nitrogren ``switch'' to regulate rigid aromatic fluorophores, pyrene and anthracene.

\begin{figure}[t]
\begin{centering}
\includegraphics[width=0.7\columnwidth]{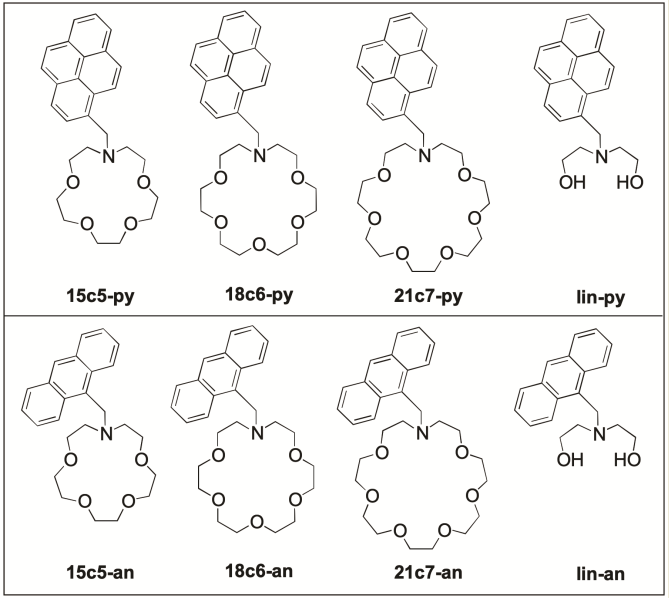}
\par\end{centering}
\caption{Molecules developed for dry barium sensing in this work. \label{fig:Molecules}}
\end{figure}

In this paper we present both wet and dry-phase barium-induced fluorescence in eight molecules constructed from four nitrogen-containing binding sites: 1-aza-15-crown-5 (15c5), 1-aza-18-crown-6 (18c6), 1-aza-21-crown-7 (21c7), and a diethanolamine linear fragment of the aza-crown ethers (lin); and two fluorophores: pyrene (py), anthracene (an).  The molecules developed for this work are shown in Fig.~\ref{fig:Molecules}.  

This paper is organized as follows.  In Sec.~\ref{sec:Wet} we report on the performance of each species in solution. We compare the response at fixed barium concentration to establish relative performance, and then undertake titration studies to establish $K_d$ values for the more promising anthracene derivatives.  In Sec.~\ref{sec:Dry} we report spectroscopic measurements of pre-chelated and dried samples to check whether fluorescence is maintained in the dry phase. In Sec.~\ref{sec:Microscope} we study these samples under a fluorescent microscope, replicating the optical configuration that will eventually be used for single molecule studies. Finally in Sec.~\ref{sec:Conclusions} we present our primary conclusions, and outline the next steps in this program.

\section{Characterization of barium sensors in solution\label{sec:Wet}}

\begin{figure}[t]
\begin{centering}
\includegraphics[width=0.99\columnwidth]{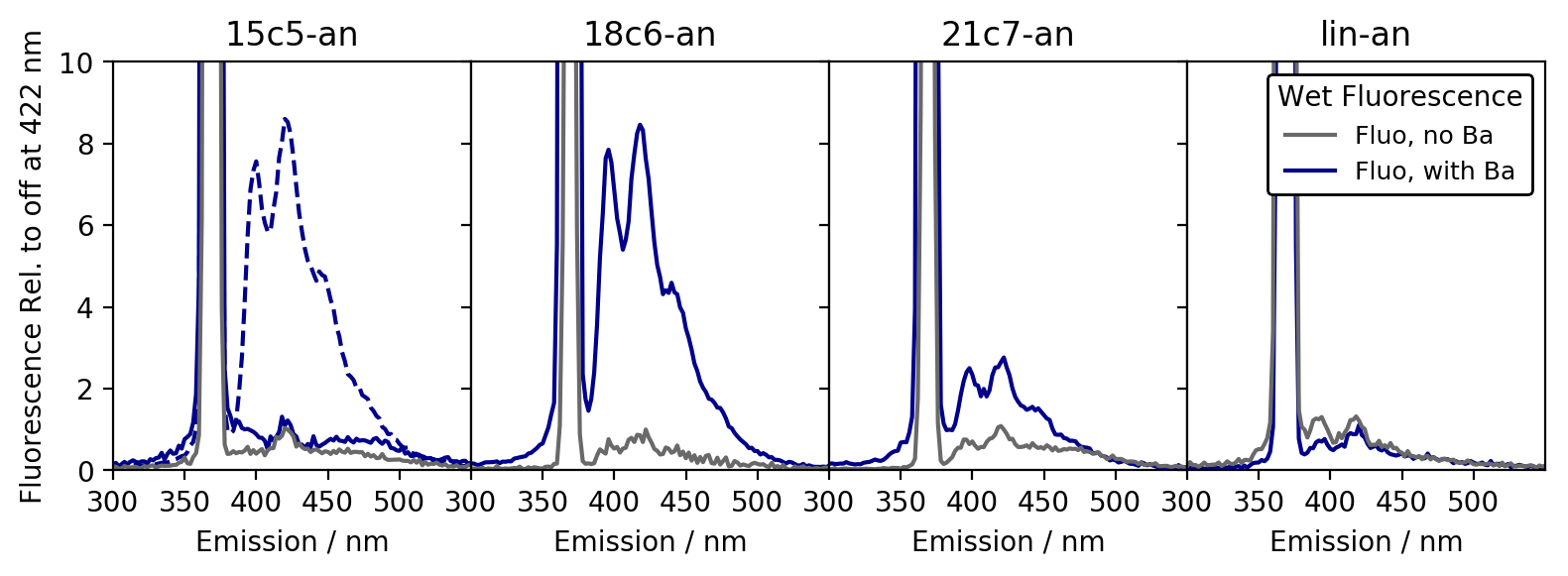}
\includegraphics[width=0.99\columnwidth]{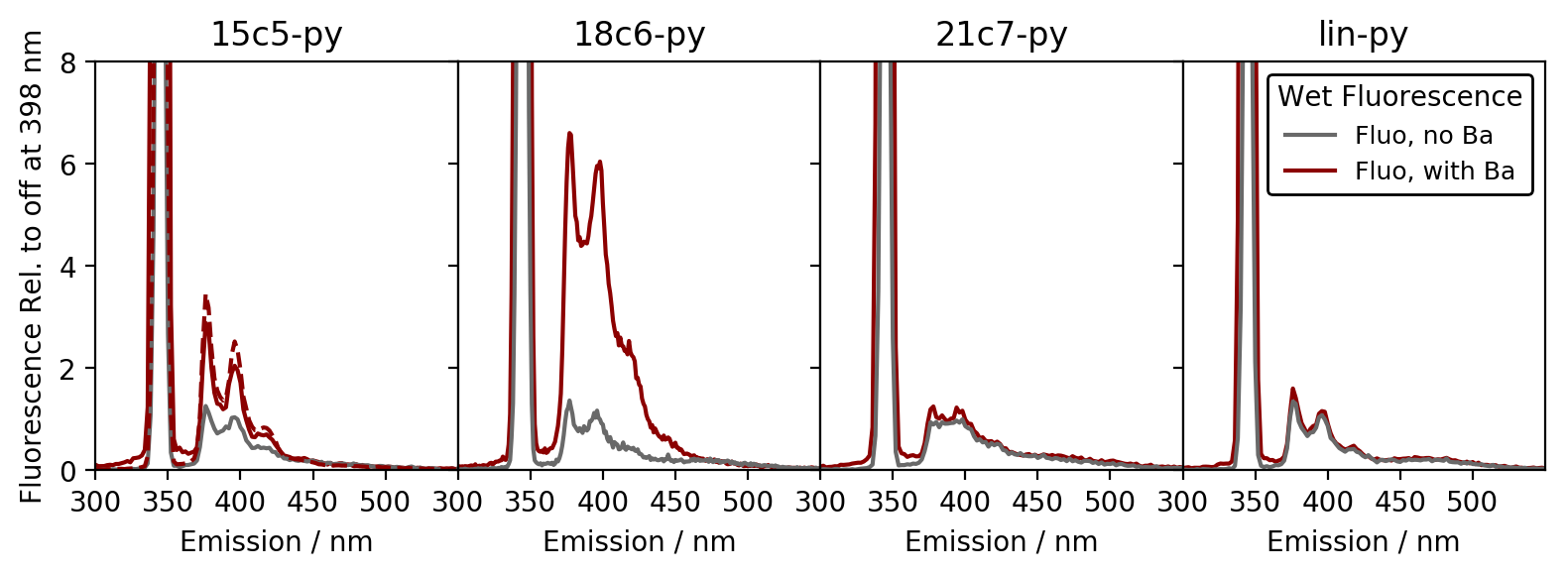}
\par\end{centering}
\caption{Fluorescence response of each species in solution at concentration of 2 $\mu$M and barium at 37.5 mM.  In each panel the grey line shows the emission contribution from scattering and fluorescence in the buffer alone.  The coloured lines show the response with barium added. The dashed line shows a higher concentration study for comparison, with the 15c5 fluorophores at 90 $\mu$M. \label{fig:WetPlots}}
\end{figure}

The fluorescence intensity of unbound and barium-bound fluorophores were first studied in solution (``wet studies'').  In order to achieve solubility of the pyrene and anthracene derivatives, a solvent mixture of 10:1 TRIS buffer and acetonitrile mixture was used.  Protonation of the nitrogen ``switch'' was observed to lead to a non-quiet off state in neutral, un-buffered solution, and so the buffer was prepared to pH 10.1 to enhance the on-off ratio. Some residual activity in the off state was still observed even at this higher pH. Data that will be shown in Sec.~\ref{sec:Dry} are suggestive that the residual off-state fluorescence is largely induced by solvent effects which are not present in the dry phase.

\begin{figure}[t]
\begin{centering}
\includegraphics[width=0.99\columnwidth]{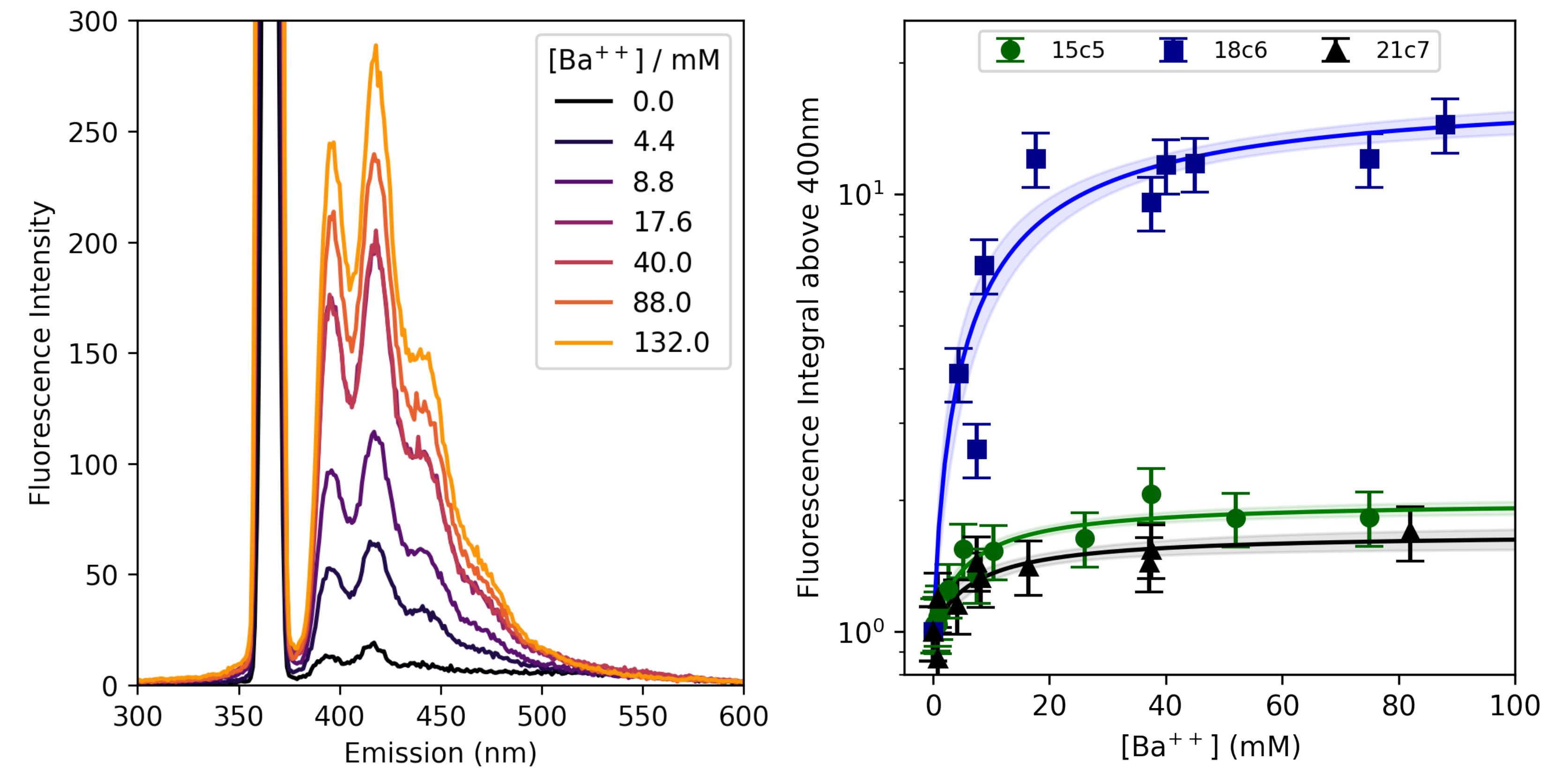}
\par\end{centering}
\caption{Left: Titration study showing increase in fluorescence intensity with added barium to 18c6-an.  Right: Titration curves for anthracene derivatives with fits for $K_d$ overlaid. \label{fig:titrationplots}}
\end{figure}

The intensity of fluorescence was measured as a function of wavelength between 300 and 550~nm.  For each fluorophores a 2D emission / excitation spectrum was recorded to establish the most efficient excitation wavelength $\lambda_{max}$, with an example for 18c6 shown in Supplemental Fig.~\ref{fig:18c6in2d}.  Experimentally we found that  $\lambda_{max}\sim$342-344 nm for the pyrene derivatives and $\lambda_{max}\sim$366-368 nm for the anthracene derivatives.  To avoid quenching and excimer formation effects observed at higher concentrations, relatively weak solutions of 2$\mu$M were used.   These concentrations are lower than preliminary results reported in~\cite{Byrnes:2019jxr}.  The various species were first compared at fixed barium concentration of 37.5 mM to establish relative sensitivity.   These data are shown as solid lines in Fig.~\ref{fig:WetPlots}, where the curves are normalized to the intensity of the ``off'' state at the longest wavelength emission peak (as indicated on the y axes).

The linear systems are not expected to capture barium in solution, but were included as control systems for wet and dry fluorescence.  As expected, we observed no significant increase in fluorescence from these species on addition of barium ions.  The strongest fluorescence response was found for the 18c6 species, which is consistent with expectations based on size-matching of the ion and the receptor site~\cite{steed2001first}. 

A smaller but still significant response is observed in some other molecules, including 21c7-an and 15c5-py.  We also found that the absence of an observable fluorescent barium response in 2 $\mu$M 15c5-an could be alleviated by working at higher concentrations of 90 $\mu m$ or above, as shown as a dashed curve in Fig.~\ref{fig:WetPlots} and in~\cite{Byrnes:2019jxr}. However, at these concentrations, efficient saturation could not be achieved without substantial excimer formation in some samples~\cite{BirksPyrene} in dry data (see Sec.~\ref{sec:Dry}), and so we continue most studies at the lower concentration point of 2 $\mu$M.

\begin{figure}[t]
\begin{centering}
\includegraphics[height=0.4\columnwidth]{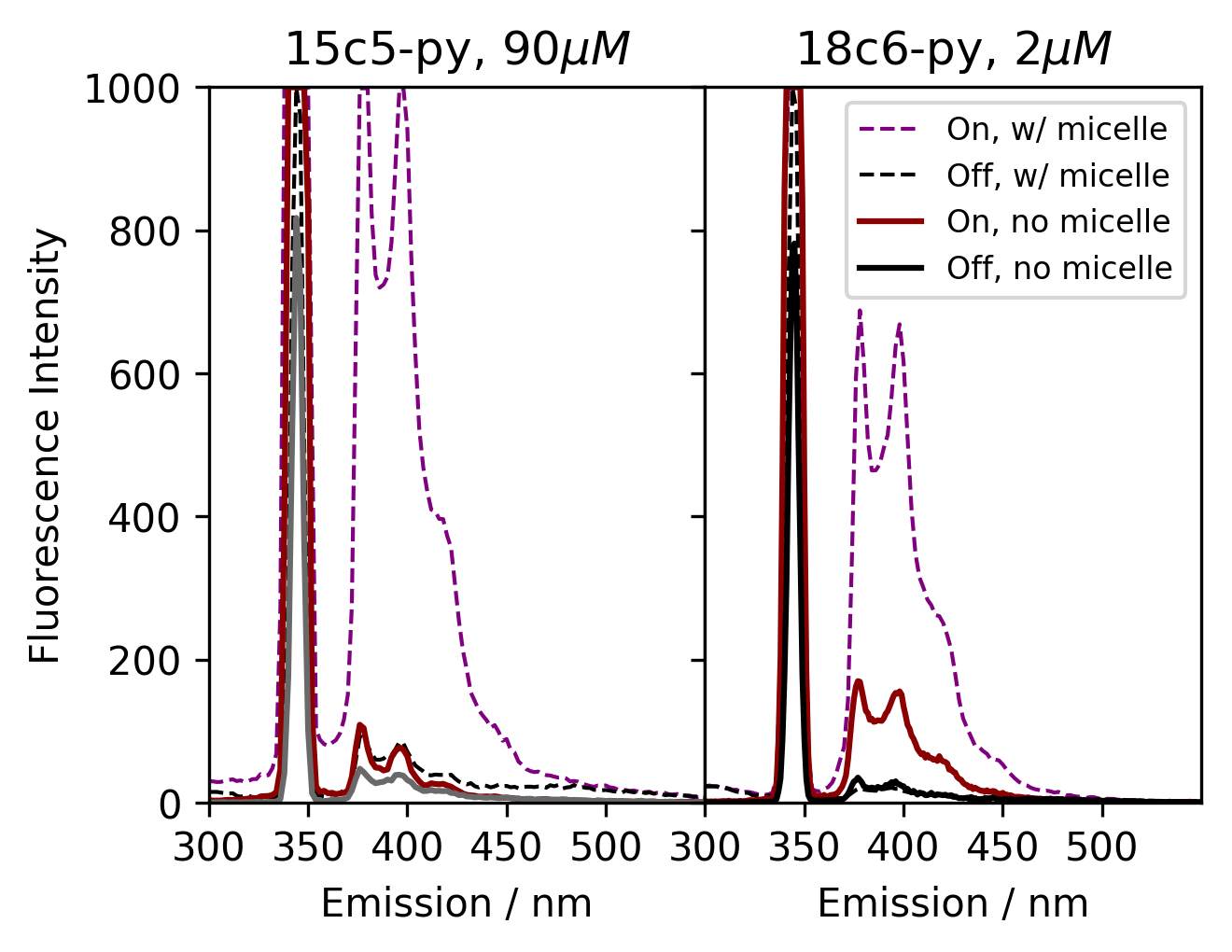}
\includegraphics[height=0.4\columnwidth]{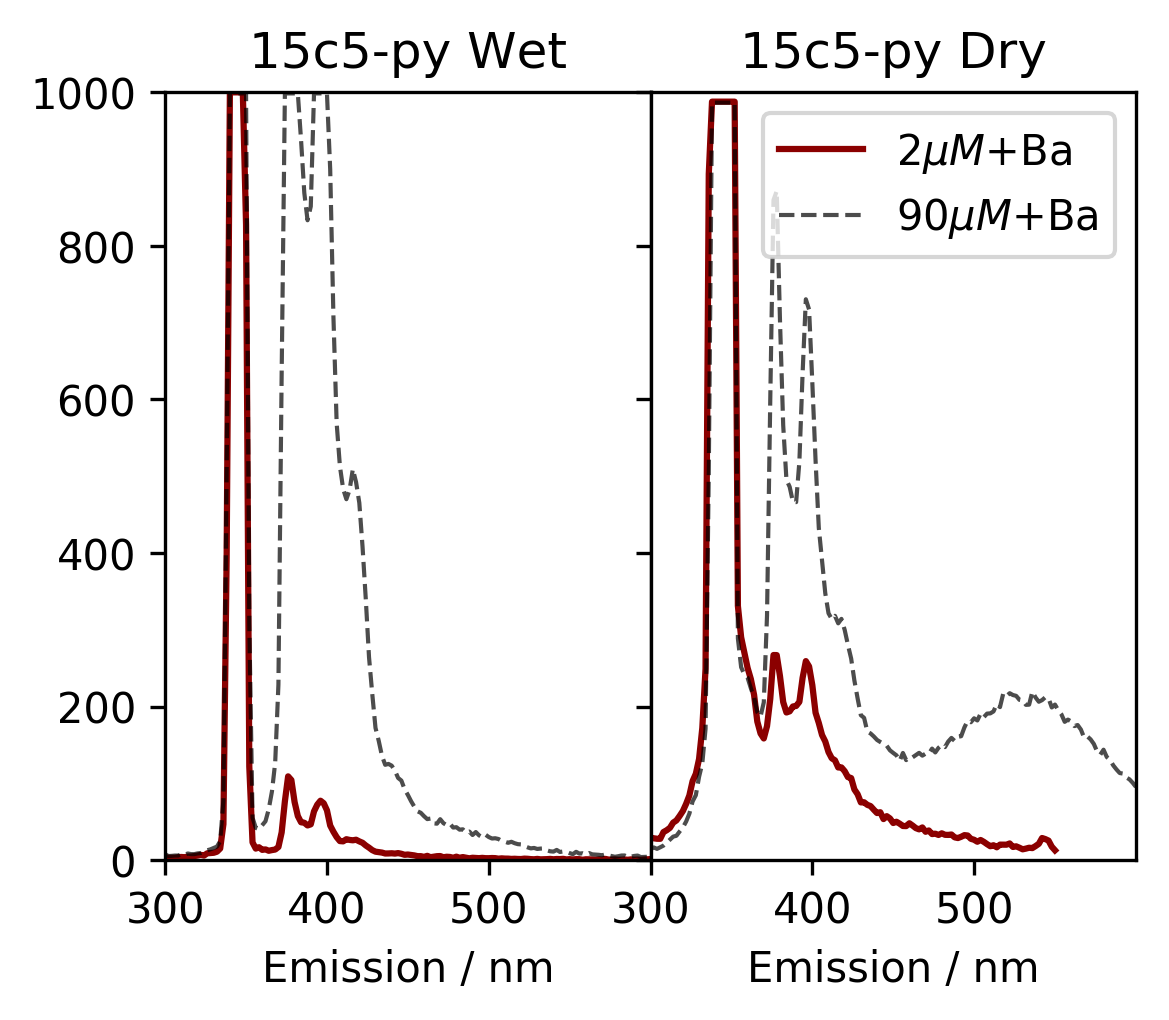}

\par\end{centering}
\caption{Left: Effect of micelles on 15c5-py and 18c6-py species, showing significant solvent effect on fluorescence in wet state. Right: effect of increasing 15c5-py concentration. A direct increase in primary fluorescence is observed in solution, whereas an excimer peak emerges in dry samples.\label{fig:Micelle}}
\end{figure}

The brightest barium response at 2$\mu$M fluorophore concentration was observed in 18c6-an, and as such, this molecule is selected as the basis for more detailed studies.  A Job's plot study demonstrated that 18c6-an binds with Ba$^{2+}$ in the ratio 1:1, shown in Supplemental Fig.~\ref{fig:JobsPlot}.  $^1$H NMR experiments of both 18c6-py and 18c6-an with or without barium perchlorate in 1:1 ratio in CD\textsubscript{3}CN suggest the involvement of nitrogen and oxygen lone pairs in interaction with barium cations as indicated by downfield shift of all proton signals in the aza-crown ether ring and the benzylic protons located between the nitrogen and anthracene. More subtle effects are observed in the aromatic region, showing a likely small field effect on the anthracene ring upon binding. Similar binding experiments with acyclic lin-an show minimal downfield shifts, indicative of less effective barium-fluorophore binding with those receptors. Two illustrative examples are shown in Supplementary Fig.~\ref{fig:NMR}.

Titration studies of barium into the anthracene derivatives are shown in Fig.~\ref{fig:titrationplots}.  For the pyrene derivatives, similar studies were undertaken, but found to be less reliable, possibly as a result of relatively slow solvation rates paired with increased rates of oxidative degradation of pyrene in our solvent system leading to larger fluctuations in a batch-to-batch manner.  The error bars in Fig.~\ref{fig:titrationplots} reflect the observed variability between repeated runs at the same concentration point.  The overlaid fit functions reflect the functional form associated with a 2-body association/dissociation reaction, with the error bands show 68\% frequentist intervals at each barium concentration, accounting for correlated variations in the value of $K_d$ and the saturation fluorescence.  The extracted $K_d$ values from the fits of Fig.~\ref{fig:titrationplots}, right are: 15c5-an $K_d=0.988\pm 0.008$ mM; 
18c6-an $K_d=16 \pm 4$ mM; 21c7-an $K_d=0.68 \pm 0.01$ mM.  The uncertainties reported there are statistical only, but some variance between repeated titrations suggests the uncertainty on each $K_d$ should be larger than this due to systematic effects.  We attribute an additional 20\% systematic uncertainty to each measured $K_d$ value on this basis, and so report 15c5-an $K_d=0.98\pm 0.20$ mM; 
18c6-an $K_d=16 \pm 5.2$ mM; 21c7-an $K_d=0.68 \pm 0.13$ mM.  The general values are reasonable, when compared to measured binding constants for all-oxygen crown ethers interacting with barium cations in water\cite{mizoue2004calorimetric}.

The fluorescence in the wet phase is strongly influenced by the solvent.  The addition of a Triton-X above its critical micelle concentration in our solvent system was studied, to produce a protective non-aqueous environment around the molecule suspended in a microemulsion, potentially more representative of the environment expected in the dry phase.  For both 15c5-py (90 $\mu$M) and 18c6-py (2 $\mu$M) solutions, the micelle substantially improved fluorescent response of the system, as shown in Fig.~\ref{fig:Micelle}, left. This offers promise that the dry response may be expected to be innately (or tailored to be) stronger than the wet response for these molecules.  For anthracene derivatives, no substantial change was observed with the addition of the micelle, which suggests that protection of the fluorophore from quenching rather than protection of the receptor from disturbance may be the dominant effect leading to improved performance for pyrene fluorophores.

\begin{figure}[t]
\begin{centering}
\includegraphics[width=0.99\columnwidth]{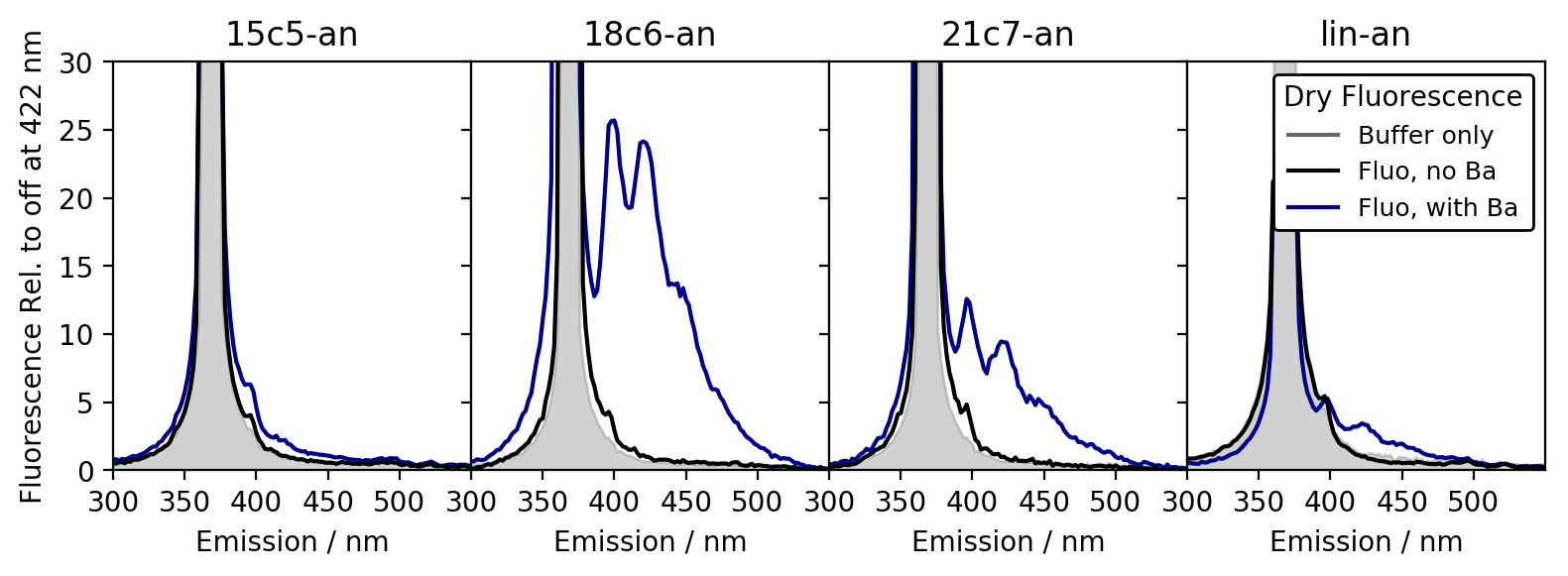}
\includegraphics[width=0.99\columnwidth]{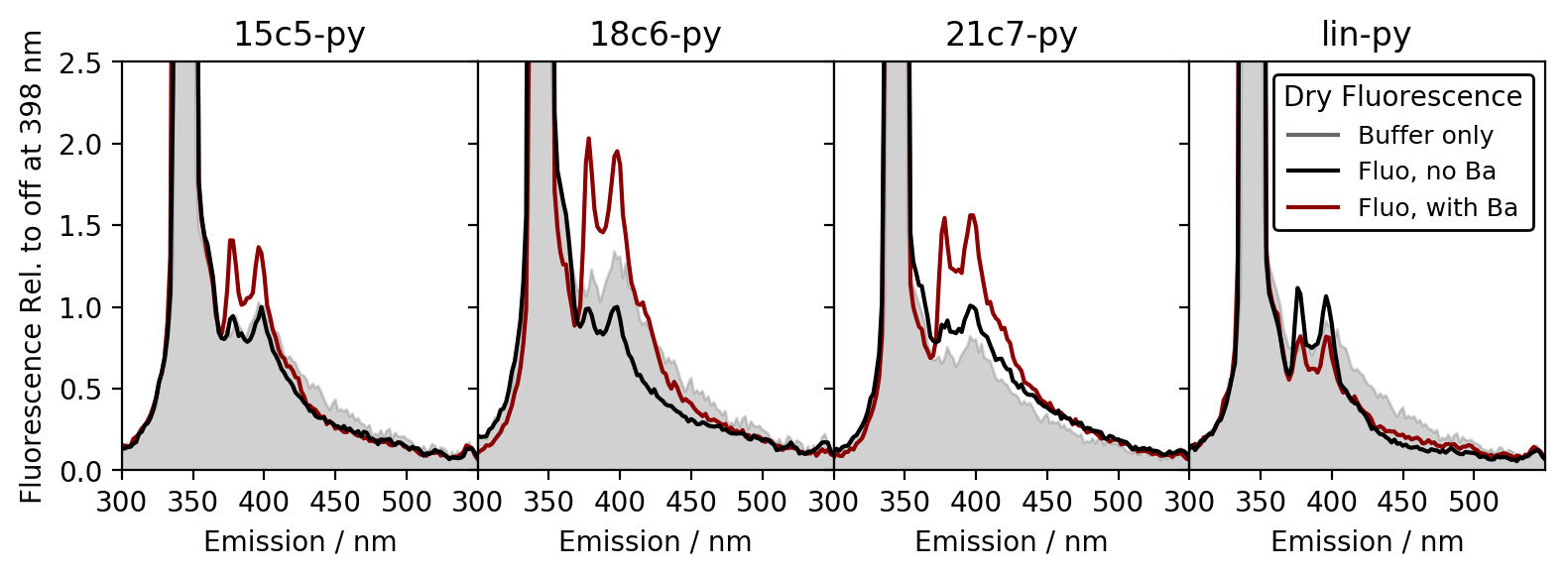}
\par\end{centering}
\caption{Fluorescence response of dried films of fluorescent barium sensing molecules in buffer.  In each panel the grey line shows the emission contribution from scattering and fluorescence in the buffer alone. The black line shows the emission contribution from un-chelated fluorophore in buffer, and the coloured lines show the response with barium added. \label{fig:DryPlots}}
\end{figure}

\section{Characterization of barium sensors in the dry-phase via spectroscopy\label{sec:Dry}}

Dry fluorescence studies of pre-chelated solutions were performed by drying 8-10 drops of each wet sample on a glass plate. The samples were dried by heating the glass on a hot plate at temperature 120 $^{\circ}$C. Two dimensional emission/excitation scans were again used to establish $\lambda_{max}$, with no significant shifts observed relative to wet studies. The fluorescence intensity measured at $\lambda_{max}$ is reported at wavelengths between 300 and 550~nm for samples without and with 37.5~mM added barium in Fig.~\ref{fig:DryPlots}.  Broadly we observe the same trends as in the wet studies, with 18c6 demonstrating the strongest response.  

In the dry phase an additional complication is present, which is that the glass and dried buffer is found to introduce some level of scattering and fluorescent response, shown in grey in Fig.~\ref{fig:DryPlots}. The effect is much larger for the pyrene species, which are excited at 344~nm, than for the anthracene derivatives, excited at 368~nm.  Fluorescence from the dried fluorophore with no barium is found to be negligibly different from the glass/buffer control system in all cases, suggesting that the isolated molecules are extremely quiet in the dry phase when unchelated.  This lends support to the hypothesis that the off-state fluorescence is a largely solvent-induced effect, caused by interaction between the nitrogen group and residual protons in the solvent.  

All anthracene derivatives show a fluorescent response to barium cations in the dry phase, with 18c6-an again being the strongest, followed by 21c7. The linear system here also shows some response, likely due to the co-precipitation of barium when rapidly solidified on the scanning surface. The pyrene derivatives show some response, with 18c6-py being strongest, but the response is more difficult to observe over the buffer / scattering background.

As in the wet studies, both 15c5 responses to barium are weak.  Following a similar strategy to the wet samples we also tested a higher concentration sample of 15c5-py in the dry phase, shown in Fig.~\ref{fig:Micelle}, right.  In this case, a large increase in fluorescence is observed in the dry phase, but at much longer wavelengths than the expected pyrene monomer emission. This emission at 525~nm derives from the pyrene excimer~\cite{BirksPyrene}, and thus represents a collective effect of fluorescence between multiple pyrene groups.  This emission is the likely origin of the intense dry fluorescence reported in~\cite{Byrnes:2019jxr}, given the optical configuration used there.  Because the excimer likley requires barium binding to two fluorophore molecules simultaneously, it is not of direct interest for the goal of single molecule imaging at a self-assembled monolayer consisting of these chemosensors.

\section{Characterization of barium sensors in the dry-phase via microscopy\label{sec:Microscope}}

\begin{figure}[t]
\begin{centering}
\includegraphics[width=0.49\columnwidth]{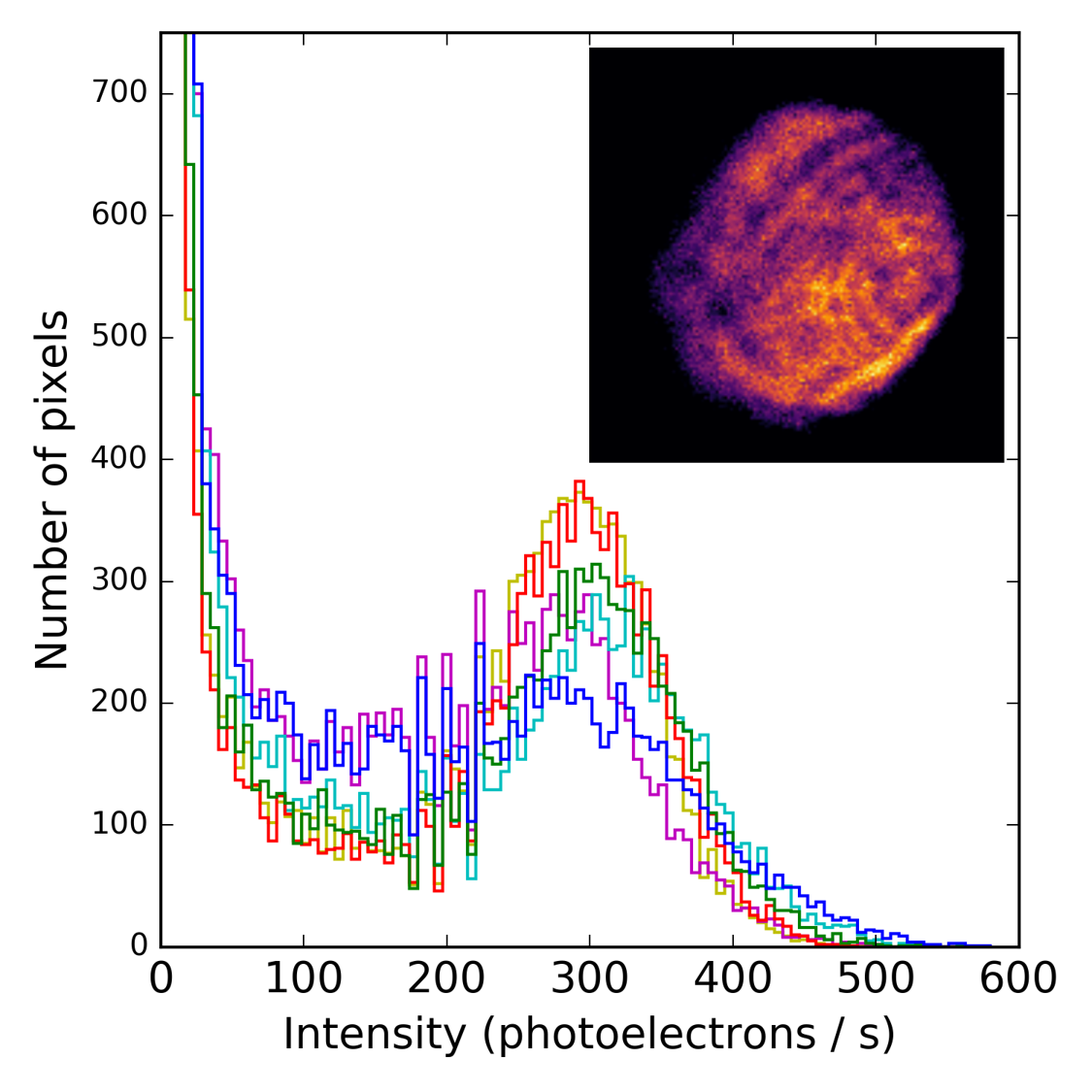}
\includegraphics[width=0.49\columnwidth]{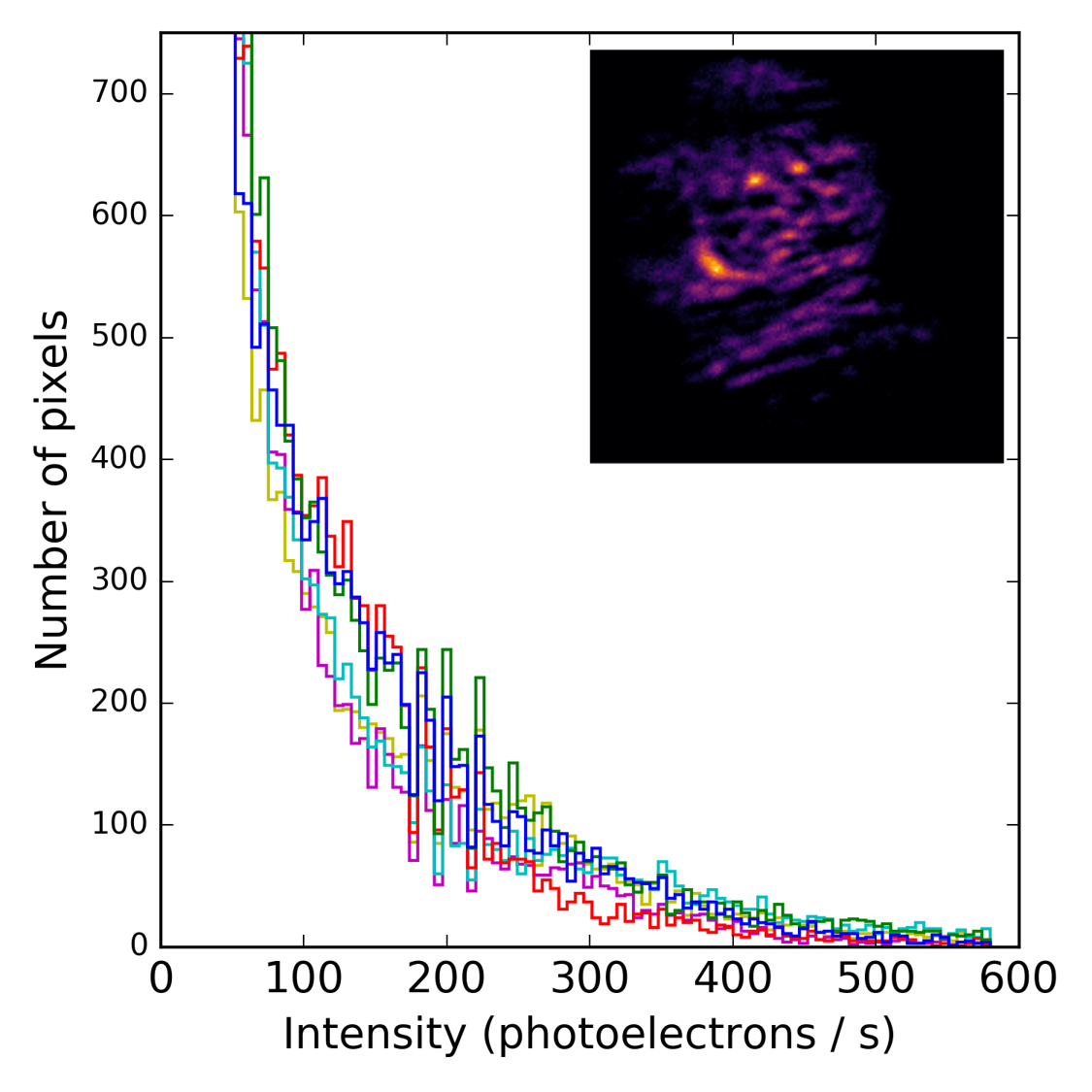}
\par\end{centering}
\caption{Example of barium-chelated (left) and unchelated (right) fluorescence microscope images. The histograms show the pixel intensity populations, with a clear excess visible in the ``on'' slide.  The multiple colored histograms are various locations within the fluorescent spot, chosen at random. The insets show two example microscope images at specific locations. \label{fig:MicroscopePlots}}
\end{figure}

\begin{figure}[t]
\begin{centering}
\includegraphics[width=0.5\columnwidth]{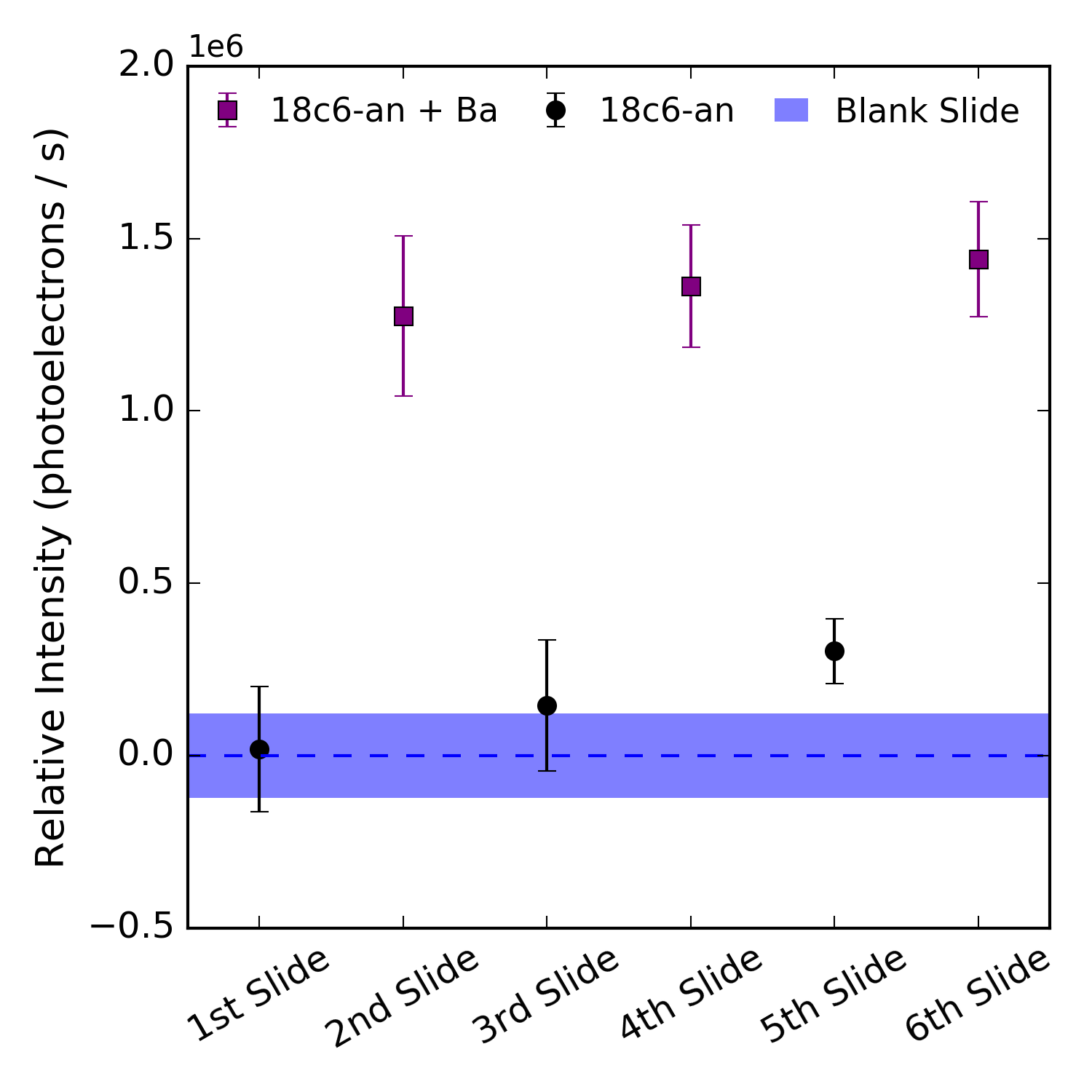}
\par\end{centering}
\caption{Measured response of dry barium-chelated vs. unchelated microscope slides excited at 365~nm under fluorescence microscope.  Each error bar represents the 1$\sigma$ spread if intensities within a single slide, and each point represents an independently prepared slide.  A clear enhancement of the ``on'' state relative to the ``off'' state is visible.\label{fig:MicroscopeHisto}}
\end{figure}

The most promising compound identified in this work is 18c6-an.  It binds strongly to barium in solution, maintains fluorescence in the dry phase, and has spectroscopic properties amenable to microscopy, with both excitation and emission at fairly long wavelengths, and a large enough stokes shift to allow dichroic separation of excitation and emission.  Samples of dried on- and off-state 18c6-an were examined under a fluorescence microscope, to establish that bulk fluorescence is detectable microscopically.

Using total internal reflection microscopy (TIRFM)~\cite{axelrod2003}, the fluorescent response of 18c6-an was measured both with and without exposure to Ba$^{2+}$, alongside blank coverslips tested as a control to monitor the level back-scattered light and camera noise.  

The slides were prepared by sonicating first for 20 minutes in a mixture of one-part Liquinox Cleaning Solution and ten parts deionized water, and then again in only deionized water for ten minutes. The slides were then left to soak in 6M Nitric Acid for four hours in order to remove any remaining metal impurities on or near the surface of the slides. The slides were then thoroughly washed with Ulta-trace water, and rinsed in buffer solution in order to neutralize any remaining acid before a final water rinse.

In order to assess the removal of residual Ba contamination from the glass slides, two tests were performed.  In one test, two sets of three cleaned and three uncleaned slides were rinsed with 2 mL of 18 MOhm deionized water, acidified with 80 mL 8M Optima Grade nitric acid to make solutions of 2\% HNO3 and analyzed via inductively couple plasma mass spectrometry (ICP-MS).  Determinations of 2.06 $\pm$ 0.77 and 18.8 $\pm$ 5.8 pg Ba were made among the three cleaned and uncleaned slides, respectively, for a reduction factor of 9.1.
 
In a second test, in individual vials, a cleaned and uncleaned slide were sonicated in 20 mL of 5\% (v/v) HNO3 in a 40 C bath for 99 minutes and allowed to soak overnight.  Determinations of 90.1 $\pm$ 0.8 and 746 $\pm$ 2 pg Ba were made from the leachates of the cleaned and uncleaned slides, respectively, for a reduction factor of 8.3.  Detection limits of 1 fg Ba were calculated (3 $\times$ standard deviation of the process blanks, n = 3). 

After allowing cleaned slides to dry, each was heated to 100 $^{\circ}$C and 37.5 $\mu$L of barium enriched and barium free 18c6-an were deposited on the glass drop-wise until dry. This forms a fluorescent spot of area around 9~mm$^2$.

The coated slides are then imaged in a system that resembles that used to image single molecules in~\cite{McDonald:2017izm}.  The optics of the system were switched to a set appropriate for anthracene fluorescence, incorporating a 450~nm short pass emission filter, 400~nm long pass emission filter, and 409~nm long-pass dichroic mirror.  A 365~nm beam is selected from a broadband super-continuum laser source and focused onto the back focal plane of an oil immersed, high numerical aperture, 100X objective lens. The light coming from the objective lens is then bent through the different refractive indices of the glass, oil, glass and air in its path, allowing the excitation of only the molecules at the surface interface. The fluorescent response for each sample from this excitation was then recorded using a Hamamatsu EM-CCD camera, at one second of exposure. This was repeated for six positions on each slide, with three slides of each type (fluorophore with no barium, fluorophore with barium, blank control slide).  Example pixel intensity histograms and images for one on-state and one off-state sample are shown in Fig.~\ref{fig:MicroscopePlots}.  The fluorescent response for the on state of 18c6-an is visibly higher overall than that for the off state in this example.  

In order to quantify the effect in aggregate, the total intensity within the region of interest above one hundred photo-electrons was summed for each point on each field of view on each slide.  The resulting fluorescence intensity measurement per slide are shown in Fig.~\ref{fig:MicroscopeHisto}. The off-state slides do not show significant activity above the blank slide baseline, but a clear increase in fluorescence in the barium-chelated samples is observed.  Some time dependent degradation of the off state was observed when exposed to air for multi-hour periods, likely due to aerobic oxidation of the SMFI species' switchable nitrogen\cite{clark2007solution}. To mitigate any biasing effect the data were taken with alternating on and off slides one after the other.  Both the clear distinction between on- and off-state fluorescence and the gradual increase with time are visible in Fig.~\ref{fig:MicroscopePlots}.  These data demonstrate that 18c6-an has appropriate spectroscopic properties for dry barium chemosensing via fluorescence microscopy.

\section{Conclusions\label{sec:Conclusions}}

We have studied a family of molecules designed for dry-phase barium chemosensing, based on azo-crown ether species bonded to small rigid dye molecules via a nitrogen switch.  Barium responsiveness shows a clear trend with 18c6 giving the strongest response, with intermediate responses observed in 15c5 and 21c7 receptors, and no in-solution response from the linear molecules, which are ineffective for binding barium.  Titration studies were undertaken, establishing $K_d$ values for all anthracene-based species. 

Dry phase fluorescence was studied spectroscopically, with 18c6 derivatives again showing the strongest response in pre-chelated and dried samples.  For all the anthracene derivatives, the off-state of the sensor molecule becomes negligible in the dry phase, with the only background deriving from residual scattering and low level fluorescence in the glass and buffer.  This is suggestive that any off-state fluorescence in the wet studies is attributable to solvent effects, likely protonation of or hydrogen-bonding to the nitrogen group preventing it from inhibiting fluorescence. Based on these studies, 18c6-an was identified as the most promising candidate for further work.

The fluorescence enhancement from barium chelation was tested under an inverted fluorescence microscope, in conditions similar to those used previously for single molecule imaging of barium ions by FLUO-3. We observe a significant increase in fluorescence for the barium-bound vs unbound samples, showing that this family of dyes has appropriate spectroscopic characteristics for imaging of barium ions in the dry phase.

Single molecule response has not been demonstrated in this work, since the aggregation and crystal structure of the dried dye prohibits observation of distinct and separated fluorescent molecules.  The next stage of this R\&D program will address this by self-assembling a monolayer containing 18c6-an onto a glass surface, presenting a well organized and dispersed fluorescent sensing layer. Based on the spectroscopic properties of 18c6-an established in this work, and its strong on-off response in the dry phase exceeding that of FLUO-3 in solution~\cite{Jones:2016qiq}, we expect to be able to resolve individual barium-bound molecules within the dry layer following barium binding, once the monolayer has been assembled.  

The path to single molecule imaging using this class of fluorophores will build upon our past work in~\cite{McDonald:2017izm}, using total internal reflection fluorescence microscopy to identify near-surface ions, first within polymer matrices and then in surface-tethered monolayers. The UV excitation of these new molecular probes presents some new challenges in terms of eliminating substrate fluorescence, which can be overcome either with the use of low-fluorescence materials or by extending the structures we have demonstrated to analogues with visible excitation and emission.  These topics will be the subject of future publications.

Demonstration of a class of molecules with a dry fluorescent barium response, observable via TIRF microscopy, is a significant advance in the development of barium tagging in high pressure xenon gas.  Since the single molecule fluorescence imaging approach is applicable directly at atmospheric or elevated pressures, these new species of barium chemosensors open a viable path to in-situ barium ion detection within a high pressure xenon gas detector.  Demonstrating performance at the single molecule level in noble gases is the next important step.  With this new family of chemosensors for dry SMFI in hand, and single ion sensitivity to barium already reported in the wet phase~\cite{McDonald:2017izm}, a near-term demonstration of single barium ion detection in dry xenon gas appears within reach.

\section*{Methods}
Commercially available chemicals were purchased from Alfa Aesar, Milipore Sigma and Acros Organics and were used without further purification, unless otherwise specified. Deionized (DI) water for ultratrace analysis was purchased from Sigma Aldrich and acetonitrile used was of LCMS grade. All reactions were performed under atmospheric conditions unless otherwise mentioned. \textsuperscript{1}H NMRs were acquired on 300 MHz and 500 MHz spectrometers and referenced to the internal solvent signals (7.26 ppm in CDCl\textsubscript{3} or 3.33 ppm in CD\textsubscript{3}OD). \textsuperscript{13}C NMRs were acquired on 75 MHz and 125 MHz spectrometers referenced to the internal solvent signals (central peak 77.00 ppm in CDCl\textsubscript{3} or 45.00 ppm in CD\textsubscript{3}OD). NMR data are reported as follows: chemical shift (in ppm, $\delta$), integration, multiplicity (s = singlet, d = doublet, t = triplet, q = quartet, m = multiplet, br = broad), coupling constant (in Hz). Thin layer chromatography was performed on silica gel coated aluminium plates (EMD Merck F254, 250 $\mu$m thickness). 254 nm ultraviolet light and ninhydrin stain (for amine containing molecules) were used to visualize spots. Flash chromatography was performed over Silicycle Silicaflash P60 silica gel (mesh 230-400) and standard grade activated Alumina (mesh 50-300).  Melting points were recorded in capillary tubes on a Mel-Temp II apparatus and were uncorrected. IR spectra were recorded in Bruker Alpha-P FT-IR Spectrometer by attenuated total reflectance on a diamond sample plate. HRMS data were recorded in Shimadzu TOF spectrometer in the Shimadzu Center for Advanced Analytical Chemistry at UT Arlington.

\begin{figure}[t]
\begin{centering}
\includegraphics[width=0.99\columnwidth]{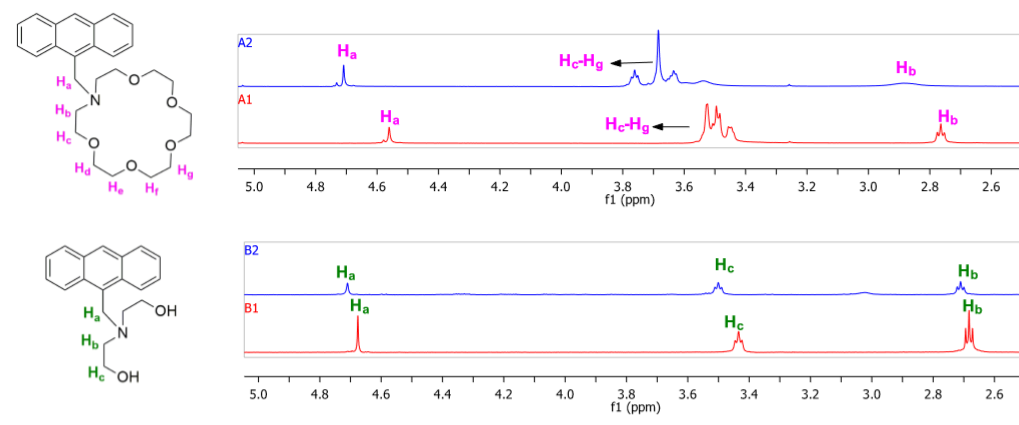}
\par\end{centering}
\caption{NMR experiment showing barium binding in 18c6-an and lin-an. Bottom (blue) traces show fluorophore without barium, and top (red) traces show fluorophore with barium in 1:1 ratio.  Lone pairs are coordinated to barium, leading to decrease in electron shielding around the proton labelled as H$_a$ in the 18c6 receptor.   The similar proton shows much smaller shift in the lin receptor, which does not have a strong tendency to bind barium. \label{fig:NMR}}
\end{figure}

\begin{figure}[t]
\begin{centering}
\includegraphics[width=0.89\columnwidth]{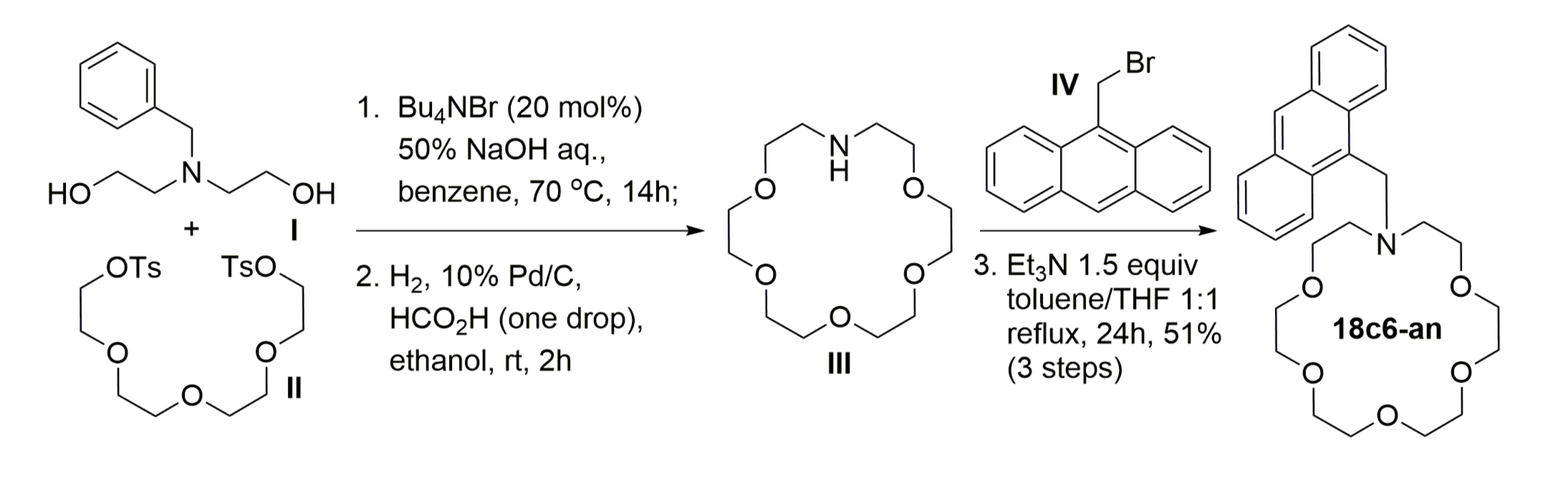}
\par\end{centering}
\caption{Example synthesis of 18c6-an. For more information see supplementary information.\label{fig:Synthesis}}
\end{figure}

\subsection{Details of fluorophore synthesis}
The synthesis of the fluorophores were carried out in a divergent method. \textbf{18c6-an's} synthesis is shown in Fig.~\ref{fig:Synthesis} as an example. (See the Supplementary Information for detailed experimental methods to all molecules studied in this report, including chemical characterization and spectra for all intermediates and final products.) Our synthesis began with the preparation of various protected diethanolamines, settling on \textit{N}-benzyldiethanolamine (\textbf{I})\cite{zhan2014integrating} to be cyclized with the activated ditosylate (OTs, \textbf{II})\cite{luk2012synthesis}. Cyclization of \textbf{I} and \textbf{II} was achieved most efficiently under a reported method that used phase transfer conditions\cite{luk2012synthesis} to yield the \textit{N}-benzyl-aza-crown ether, which was deprotected by standard hydrogenolysis to provide 1-aza-18-crown-6 ether (\textbf{III}) in good yield. 9-bromomethylanthracene (\textbf{IV})\cite{malwal2015benzosulfones} was prepared in excellent yields by radical bromination and underwent an S\textsubscript{N}2 displacement by (\textbf{III} to provide \textbf{18c6-an} in reliable yields.

\subsection{Spectrometer protocols}
\begin{figure}[t]
\begin{centering}
\includegraphics[width=0.9\columnwidth]{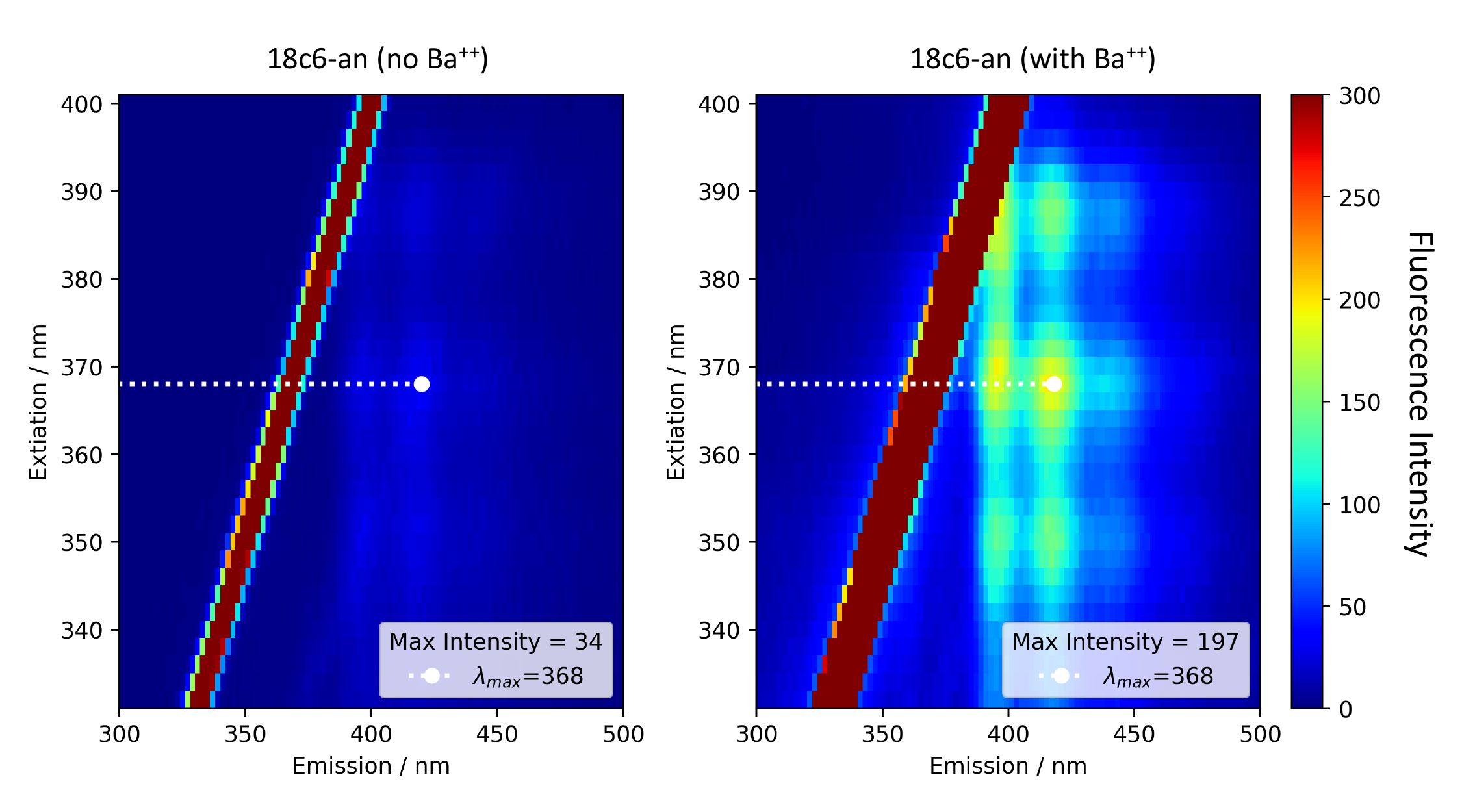}
\par\end{centering}
\caption{Fluorescence response of 18c6-an shown as a function of emission and excitation. The diagonal band represents directly reflected / scattered light \label{fig:18c6in2d}}
\end{figure}

\textbf{Instrument Set up:}
All fluorescence spectra were measured in a Cary Eclipse Fluorescence Spectrophotometer from Agilent Technologies (Product Number G8800A) at 25 $^{\circ}$C. Unless otherwise specified, instrumental set up includes: PMT detector voltage = High, Excitation filter= Auto, Emission filter = Open, Excitation slit width = 5 nm, Emission slit width = 5 nm, scan control = medium, spectral range = 300 nm to 600 nm. For each species the $\lambda_{max}$ value was first established via a two-dimensional excitation / emission scan (for example, Fig.~\ref{fig:18c6in2d}) to establish the excitaiton wavelength for subsequent tests.
\subsection{Fluorescence Samples}
TRIS [tris(hydroxymethyl)aminomethane] buffer was prepared at 20 mM concentration and the pH of the solution was adjusted to 10.1 by titrating with 1M NaOH. Stock solutions for fluorescence studies were prepared by dissolving the appropriate sample in 10:1 TRIS buffer and acetonitrile mixture using sonication and vortex instrument for total of 30 minutes. Different fluorescence samples in wet studies were prepared at concentration of 2 $\mu$M by diluting the appropriate volume of stock solution with TRIS buffer. Stock solution of barium (150 mM) was prepared by dissolving BaClO\textsubscript{4} salt in TRIS buffer. Samples were incubated for five minutes before reading the fluorescence intensity.

\begin{figure}[t]
\begin{centering}
\includegraphics[width=0.49\columnwidth]{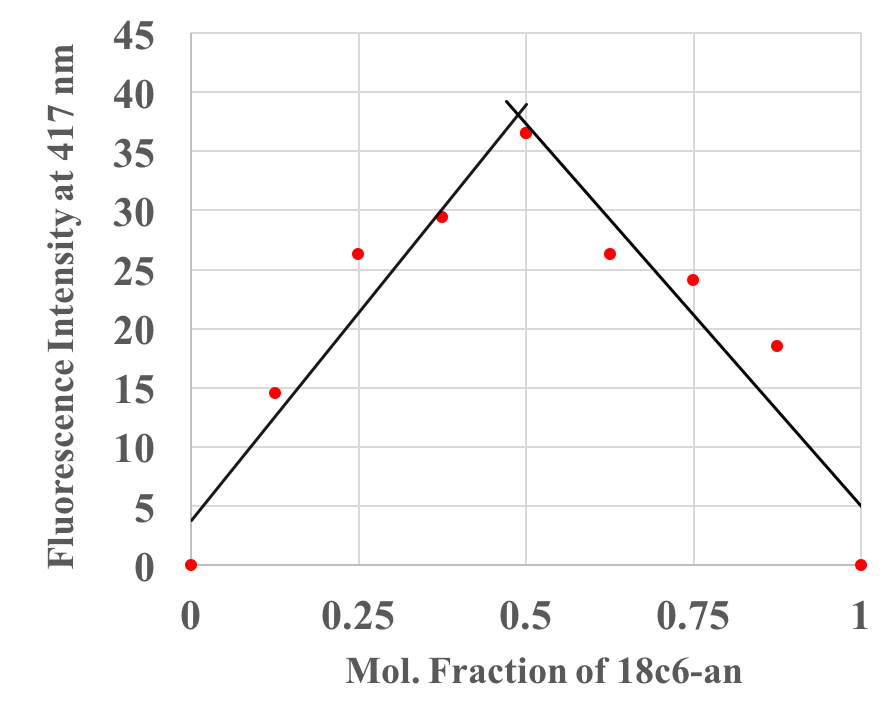}
\par\end{centering}
\caption{Job’s plot of 18c6-an and Ba$^{2+}$ with total concentration of 40 $\mu$M, which demonstrates a 1:1 binding stoichiometry \label{fig:JobsPlot}}
\end{figure}

\textbf{Job's plot study:}
Each sample for Job's plot was prepared by maintaining the total concentration of barium solution and  \textbf{18c6-an} sample at 40 $\mu$M (Fig.~\ref{fig:JobsPlot}). Fluorescence intensity was measured by obtaining the difference in fluorescence intensity value at 417 nm between sample with and without Barium. Fluorescence intensity at 1 mole fraction of 18C6-anthracene sample was normalized to 0.

\textbf{Critical Micelle concentration study:}
Critical micelle concentration was measured by obtaining the I\textsubscript{complex}/ I\textsubscript{free} values against varying concentration of Triton-X. The concentration of 18C6-py sample and barium perchlorate solution was fixed at 2.1 $\mu$M and 7.5 mM respectively for each sample analyzed. I\textsubscript{complex} refers to fluorescence intensity of solution of Triton-X, 18C6-py and barium and I\textsubscript{free} refers to the fluorescence intensity of solution of Triton-X and 18C6-py without Barium. Critical micelle concentration was obtained using linear squares fitting method for fluorescence emission at 376 nm during excitation at 342 nm against increasing concentrations of Triton-X.

\bibliography{main}

\section*{Acknowledgements}

We thank Jacqueline Baeza-Rubio for proof reading this manuscript and for her work on ion delivery, Denise Huerta for contributions to understanding gas-phase ion transport in membranes, and Charleston Newhouse and Branston Mefferd for support of synthetic and fluorescence efforts. We acknowledge the NEXT collaboration for their support and feedback.

BJPJ, AD and ADM are supported by the Department of Energy under Early Career Award number DE-SC0019054. The University of Texas at Arlington NEXT group is also supported by Department of Energy Award DE-SC0019223. 

\section*{Author contributions statement}

Molecular synthesis and spectroscopic studies were undertaken by Thapa and Denisenko. Byrnes undertook the microscopy studies.  Thapa is supported under start-up funds from BJPJ;  Denisenko by partial research assistantship from DE-SC0019054 and partial teaching assistantship, and Byrnes by teaching assistantship for this work.  Jones and Foss supervised and contributed to data analysis and presentation. Arnquist developed and validated the barium leeching procedure with ICPMS with support from DOE at PNNL.  Nygren originated the concept of SMFI-based barium tagging.  McDonald and Woodruff participated in an advisory capacity.

\section*{Additional information}

The authors declare no competing interests.

\end{document}